\documentclass[]{aastex7}

\newcommand\degd{\ifmmode^{\circ}\!\!\!.\,\else$^{\circ}\!\!\!.\,$\fi}

\newcommand{\msun}{{\rm\ M_{\sun}}}

\newcommand{\lsim}{\stackrel{\scriptstyle <}{\scriptstyle \sim}}
\newcommand{\gsim}{\stackrel{\scriptstyle >}{\scriptstyle \sim}}
\newcommand{\psr}{PSR J1745--2900}
\newcommand{\sgra}{Sgr~A*}
\newcommand{\kms}{\ensuremath{{\rm km\,s}^{-1}}}
\newcommand{\masy}{\ensuremath{{\rm mas\,yr}^{-1}}}

\begin{document}

\title{Long-Term Astrometric Monitoring of the Galactic Center Magnetar \psr}

\author[0000-0003-4056-9982]{Geoffrey C. Bower}
\affiliation{Academia Sinica Institute of Astronomy and Astrophysics, 645 N. A'ohoku Pl., Hilo, HI 96720, USA}
\affiliation{Department of Physics and Astronomy, University of Hawaii at Manoa, 2505 Correa Road, Honolulu, HI 96822, USA}
\email{gbower@asiaa.sinica.edu.tw}
\author[0000-0001-9434-3837]{Adam~T.~Deller}
\affiliation{Swinburne University of Technology}
\email{}
\author[0000-0002-6664-965X]{Paul B.\ Demorest}
\affiliation{National Radio Astronomy Observatory, P.O. Box O, Socorro, NM 87801, USA}
\email{}
\author[0000-0003-3903-0373]{Jason Dexter}
\affiliation{JILA and Department of Astrophysical and Planetary Sciences, University of Colorado, Boulder, CO 80309, USA}
\email{}
\author[0000-0003-4468-761X]{Andreas Brunthaler}
\affiliation{Max-Planck-Institut für Radioastronomie, Auf dem Hügel 69, D-53121 Bonn, Germany}
\email{}
\author[0000-0003-3922-4055]{Gregory Desvignes}
\affiliation{Max-Planck-Institut für Radioastronomie, Auf dem Hügel 69, D-53121 Bonn, Germany}
\affiliation{LESIA, Observatoire de Paris, Université PSL, CNRS, Sorbonne Université, Université de Paris, 5 place Jules Janssen, F-92195 Meudon, France}
\email{}
\author[0000-0001-6196-4135]{Ralph P. Eatough}
\affiliation{National Astronomical Observatories, Chinese Academy of Sciences, 20A Datun Road, Chaoyang District, Beijing 100101, PR China}
\affiliation{Max-Planck-Institut für Radioastronomie, Auf dem Hügel 69, D-53121 Bonn, Germany}
\email{}
\author[0000-0002-2526-6724]{Heino Falcke}
\affiliation{Department of Astrophysics, Institute for Mathematics, Astrophysics and Particle Physics (IMAPP), Radboud University, P.O. Box 9010, 6500 GL Nijmegen, The Netherlands}
\email{}
\author[0000-0002-2542-7743]{Ciriaco Goddi}
\affil{
Universidade de São Paulo, Instituto de Astronomia, Geofísica e Ciências Atmosféricas, Departamento de Astronomia, São Paulo, SP 05508-090, Brazil}
\affiliation{Dipartimento di Fisica, Universit\'a degli Studi di Cagliari, SP Monserrato-Sestu km 0.7, I-09042 Monserrato,  Italy}
\affiliation{INAF - Osservatorio Astronomico di Cagliari, via della Scienza 5, I-09047 Selargius (CA), Italy}
\affil{
INFN, Sezione di Cagliari, Cittadella Univ., I-09042 Monserrato (CA), Italy}
\email{}
\author[0000-0002-4175-2271]{Michael Kramer}
\affiliation{Max-Planck-Institut für Radioastronomie, Auf dem Hügel 69, D-53121 Bonn, Germany}
\email{}
\author[0000-0001-8403-8548]{F. Yusef-Zadeh} 
\affiliation{Dept Physics and Astronomy, CIERA, Northwestern University, 2145 Sheridan Road, Evanston , IL 60207, USA}
\email{}

\begin{abstract}
We present new astrometric observations of the Galactic Center magnetar, \psr, with the Very Long Baseline Array (VLBA).  Combined with previously published measurements in 10 epochs that spanned 477 days, the complete data set consists of 25 epochs and 41 independent measurements that span 1984 days.  These data constrain the proper motion to an accuracy of $\lesssim 2\%$ and set an upper limit on the absolute value of the magnetar's acceleration 
of $\lesssim (0.4, 0.2)\, {\rm mas\,y^{-2}}$ in the two celestial coordinates, consistent with the maximum value of $\sim 0.03\,{\rm mas\,y^{-2}}$ expected for an orbit around \sgra. Future measurements have the potential to detect the acceleration of \psr\ due to \sgra\ should \psr\ re-brighten.
We consider several potential sources of systematic variations in the astrometric residuals after fitting for standard parameters, including refractive wander, changes in the structure of \sgra, and the presence of an unseen binary companion. While a stellar companion model can be fit to the astrometric data, pulse period measurements are inconsistent with that model. No changes in the apparent image size of the magnetar were detected over the duration of these observations, indicating a lack of change in the properties of the line-of-sight scattering during this period.   We also show that the upper limit to the mean core shift of \sgra\ is consistent with expectations for a compact jet or symmetric accretion flow. 
\end{abstract}

\section{Introduction}

Pulsars in the Galactic Center (GC) are powerful tools to probe the interstellar medium, star formation and stellar death in extreme environments, and general relativity.  A pulsar in a close orbit around 
\sgra, the $4\times 10^6 \msun$ black hole in the GC, would provide unprecedented accuracy in measurement of the black hole mass and spin as well as test non-standard theories of gravity and black holes 
\citep{2012ApJ...747....1L,2016ApJ...818..121P}.  Timing of such a pulsar would strongly complement measurements from near-infrared (NIR) stellar astrometry \citep{2019A&A...625L..10G,2019Sci...365..664D} and  black hole imaging by the Event Horizon Telescope \citep{2022ApJ...930L..12E}.

A large number of GC pulsars has been predicted as a result of the known populations of high mass stars in the central parsecs \citep{2004ApJ...615..253P}.  Diffuse gamma-ray emission has also been interpreted as the result of a population of millisecond pulsars \citep{2013PhRvD..88h3521G}.  Extensive searches have been conducted at a wide range of radio wavelengths, however, revealing only a small number of confirmed pulsars in the central tens of parsecs \citep{2006MNRAS.373L...6J,2009ApJ...702L.177D,2010ApJ...715..939M,2021MNRAS.507.5053E,2023ApJ...959...14T,2024ApJ...975...34F},  \added{including the recently-discovered millisecond pulsar J1744-2946 at a distance of $\sim 1$ deg from \sgra \citep{2024ApJ...967L..16L}}.  
Deep X-ray observations of the GC also have failed to detect a large number of pulsar candidates \citep{2009ApJS..181..110M}.

The absence of radio wavelength detections has primarily been explained as the result of the effects of strong interstellar scattering in the GC, which can smear the pulsed radio emission to a timescale longer than the pulse period.  The interstellar scattering properties are primarily determined by the angular broadening of \sgra \citep{2006ApJ...648L.127B,2018ApJ...865..104J}, with supporting evidence from the angular broadening of stellar OH masers \citep{1992ApJ...396..686V}, counts of extragalactic background sources \citep{1998ApJS..118..201L,1998ApJ...505..715L},  and other measurements.  High frequency radio observations have been used as a means to mitigate the strong frequency-dependence of temporal broadening.  However, typical pulsar spectra also decrease steeply with increasing frequency, reducing the number of detectable objects.
Future searches with the SKA and with the ngVLA may be able to overcome these combined challenges \citep{2018ASPC..517..793B}.

The GC magnetar, \psr, was discovered in 2013 through a luminous X-ray burst \citep{2013ApJ...770L..24K,2013ApJ...770L..23M} that was soon followed by discovery of radio pulsations \citep{2013Natur.501..391E}.  Given the projected distance of $\sim 0.1$ pc from \sgra, its large dispersion measure and rotation measure, the high X-ray column density, \psr\ appears to be the first pulsar known to be in orbit around \sgra, with a period $\gsim 700$ years \added{\citep{Bower15}}.  Astrometric observations showed that the pulsar has a proper motion consistent with originating from  the clockwise (CW) stellar disk, which includes many O and B type stars that could serve as progenitors for a neutron star or magnetar \citep{Bower15}.  The magnetar has mostly faded in X-rays and radio emission in the years since its initial discovery \citep[e.g.,][]{2014ApJ...786...84K,2015ApJ...811L..35Y,Eatoughetal}.

The low radio frequency detection of \psr\ and a combination of the temporal and angular broadening properties has revised our understanding of the scattering properties towards \psr\ and \sgra\ 
\citep{2014ApJ...780L...2B,2014ApJ...780L...3S}.  The measured angular broadening towards \psr\ is consistent with that of \sgra, while the temporal broadening is much less than predicted if the scattering arises in the GC region.  Rather, the scattering appears to originate from a screen situated at a distance of $\Delta\sim 5$ kpc from the GC, which cannot be responsible for the absence of other pulsars as the result of temporal broadening.  Patchiness in the ISM (where other sightlines do experience heavier temporal broadening)  has been invoked as a possible geometry that would still lead to obscuration of a significant fraction of the GC pulsar population \citep{2014ApJ...780L...3S}.  Studies of other GC pulsars suggest that scattering towards these sources arises at a range of distances along the line of sight \citep{2017MNRAS.471.3563D}.  Alternatively, the extreme densities and magnetic field strengths of the GC may lead to other paths for star formation that favor creation of short-lived magnetars over ordinary pulsars \citep{2014ApJ...783L...7D}.  Detailed modeling that incorporate search sensitivities are necessary to constrain the GC pulsar population \citep{2014MNRAS.440L..86C}.

VLBI imaging and astrometry of \psr\, have been critical for addressing the source of interstellar scattering and for demonstrating the likely origin of the magnetar in the CW stellar disk.  Continued VLBI imaging probes the structure of the scattering medium as the pulsar and the Sun move relative to the scattering medium (or mediums).  Continued VLBI astrometry has the power to detect the acceleration of the magnetar due to the gravity of \sgra, confirming that it is in a bound orbit.  Additionally, astrometry can probe wavelength- and temporal-dependent 
changes in the structure of \sgra, which serves as the astrometric reference source.  

In this paper, we extend the results first 
reported in \citet{2014ApJ...780L...2B} and \citet{Bower15} with 
new observations from the Very Long Baseline Array (VLBA).  Section 2 presents the observations and data reduction.  Section 3 presents the astrometric analysis.  Section 4 presents the results on astrometry, scattering, and \sgra, incorporating new timing measurements from \citet{Eatoughetal}.  Section 5 summarizes the conclusions.

\section{Observations and Data Reduction}

\citet{Bower15} reported 15 independent position measurements in 10 epochs made with the Very Long Baseline Array (VLBA) over 477 days, beginning at MJD 56422.  New observations were obtained over the MJD range 57073 - 58406 with 26 independent position measurements in 15 epochs.  The total span of the complete data set is 1984 days with 41 independent position measurements in 25 epochs.  Observations used all available VLBA antennas, sometimes including a single ``Y1" antenna from the Very Large Array (VLA) that provided additional short baselines.  Receivers were tuned to several different frequencies, represented by the bands X (8.6 GHz, 3.4 cm), Ku (15.3 GHz, 2.0 cm), K (21.9 GHz, 1.3 cm), and Q (43.1 GHz, 0.7 cm).  Data were recorded with bit rates of 2048 Mb/s from each station.  Each observation was correlated three times: once at the location of \sgra, and twice at the location of \psr.  In the first correlation centred at \psr, pulsar gating was applied using the capabilities of the DiFX software correlator  \citep{Deller07} to retain only the $\sim$5\% of the pulse period that contained significant pulsed emission, while the second generated a standard continuum visibility dataset.  An integration time of 3.76355 seconds was used, meaning that one pulse would be contained in each integration. Using the tools described in \citet{Bower15}, a scaled version of the continuum visibilities was subtracted from the gated dataset, effectively removing the (time and bandwidth smeared) contribution of \sgra\ from the \psr\ visibility dataset and significantly improving the fidelity of the final images.

Data analysis was performed using AIPS following the same procedures in \citet{Bower15}. Primary steps in this analysis include amplitude calibration, instrumental and atmospheric delay and phase calibration on \sgra\ and transfer of solutions to the magnetar.   Delay, rate, and gain solutions on \sgra\ were derived using the same (observing band dependent) models used in \citet{Bower15}, whose apparent, angular-broadened size as a function of wavelength are consistent with past measurements \citep{2006ApJ...648L.127B}. Astrometric results are obtained from fitting in the image domain and reported in Table~\ref{tab:results}. As a result of this calibration scheme, the fundamental measurement of this experiment is the position of the magnetar relative to the position of \sgra.  

\begin{figure}[htb]
\includegraphics{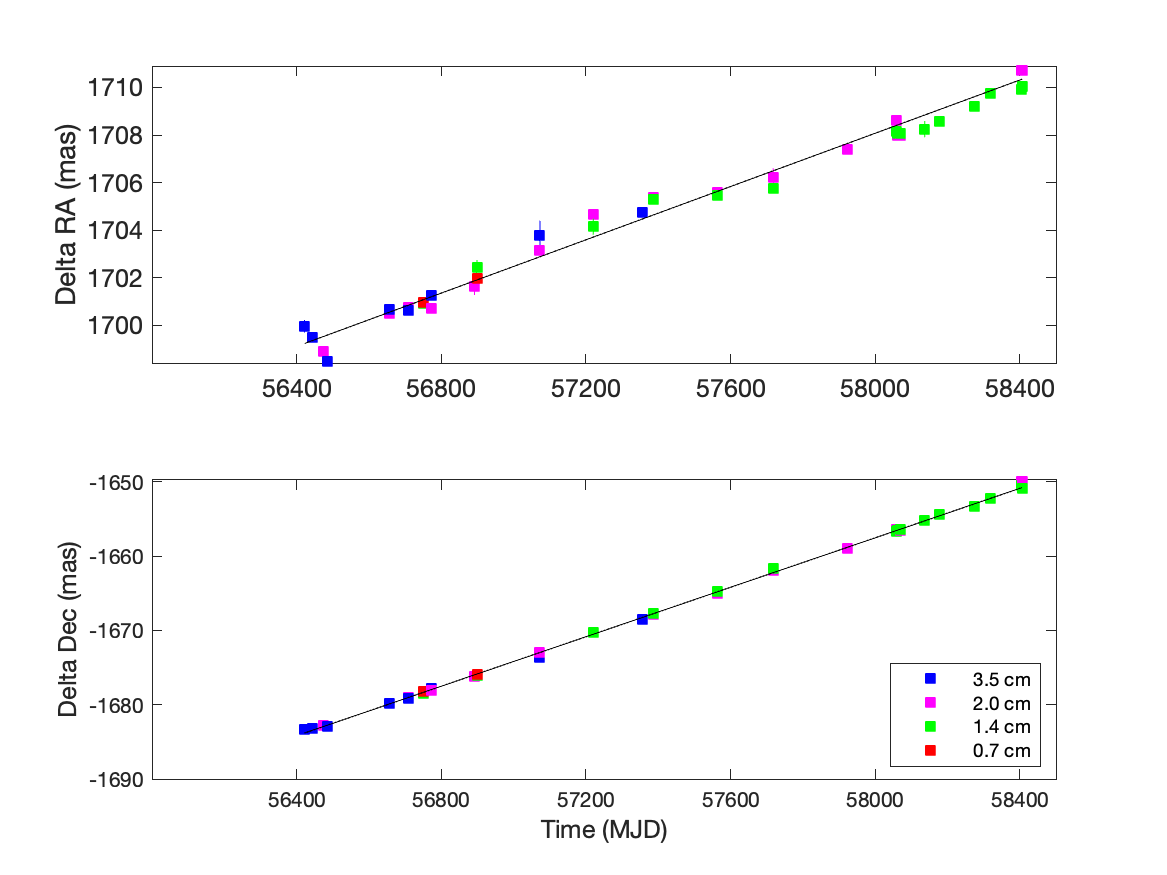}

\caption{Position as a function of time for \psr\ relative to \sgra.  Different colors are used to identify the
wavelength of observations.  The solid black line shows the best-fit proper motion in each coordinate with $\mu_\alpha=2.04 \masy$ and $\mu_\delta=6.07 \masy$.  Fits including acceleration and core shift are indistuinghisable on this scale.
\label{fig:positions}
}
\end{figure}

In \citet{Bower15}, we determined ICRF positions for the magnetar using the astrometry from \citet{2004ApJ...616..872R} with
a proper motion of \sgra\ is $\mu_{Sgr A*}=(-3.151, -5.547)$ $\masy$ and position (17:45:40.0366,--29:00:28.217) at \added{epoch 2014.145} (MJD=56710).  In this paper, we update the reference position and proper motion based on more recent measurements.  We adopt a proper motion
$\mu_{Sgr A*}=(-3.156, -5.585)$ $\masy$ from \citet{2020ApJ...892...39R}, which is consistent within the errors of the 2004 measurement.  An improved value for the position has also been presented in \citet{2023AJ....165...49G}, which we adopt here:
(17:45:40.034047,--29:00:28.21601) at epoch 2015.0.  This position corresponds to a shift of (40.1, 8.8) mas relative to the position from \citet{2004ApJ...616..872R} at epoch 2015.0.
Coordinates in the ICRF ($\Delta\alpha$, $\Delta\delta$) are given relative to this new fiducial \sgra\ position and proper motion.  For epochs published in \citet{Bower15}, the ICRF values reported here are an update.  The relative position of the magnetar to \sgra\ as a function of time are shown in Figure~\ref{fig:positions}.

We also produce fits to the observed (scatter-broadened) size of the magnetar, based on the fitting of a two-dimensional Gaussian component to the clean image followed by deconvolution of the synthesized beam (Table~\ref{tab:sizes}). 
We note that this approach is sensitive to the modeled size of \sgra\ (as it is used as a calibration source), and furthermore that image deconvolution (and the alternative approach of visibility model-fitting) are increasingly challenging at low signal--to--noise ratios and for highly resolved sources. Accordingly, we produce these fits only for a single band (2 cm) where we have the most significant detections (S/N ratio $>30$).

\begin{deluxetable}{rlllll}
\tabletypesize{\footnotesize}
\tablecaption{Observed Positions of \psr\ \label{tab:results}}
\tablehead{
\colhead{MJD} & \colhead{Band} &  \colhead{ICRF RA} & \colhead{ICRF Dec.} & \colhead{$\Delta\alpha$} & \colhead{$\Delta\delta$}
\\
              &                &   \colhead{(J2000)} & \colhead{(J2000)} &  \colhead{(mas)} &  \colhead{(mas)}
}
\startdata
    56422 &  X & 17 45 40.164031 $ \pm  0.000020 $ & -29 00 29.8902 $\pm 00.0002 $ & $ 1699.97 \pm    0.26 $ & $ -1683.38 \pm    0.20 $ \\ 
    56444 &  X & 17 45 40.163981 $ \pm  0.000012 $ & -29 00 29.8904 $\pm 00.0001 $ & $ 1699.50 \pm    0.16 $ & $ -1683.24 \pm    0.10 $ \\ 
    56473 & Ku & 17 45 40.163917 $ \pm  0.000008 $ & -29 00 29.8904 $\pm 00.0001 $ & $ 1698.91 \pm    0.10 $ & $ -1682.80 \pm    0.10 $ \\ 
    56486 &  X & 17 45 40.163876 $ \pm  0.000008 $ & -29 00 29.8907 $\pm 00.0001 $ & $ 1698.49 \pm    0.10 $ & $ -1682.90 \pm    0.10 $ \\ 
    56658 & Ku & 17 45 40.163918 $ \pm  0.000007 $ & -29 00 29.8902 $\pm 00.0001 $ & $ 1700.52 \pm    0.09 $ & $ -1679.79 \pm    0.10 $ \\ 
    \dots &  X & 17 45 40.163930 $ \pm  0.000010 $ & -29 00 29.8902 $\pm 00.0001 $ & $ 1700.68 \pm    0.13 $ & $ -1679.79 \pm    0.10 $ \\ 
    56710 & Ku & 17 45 40.163903 $ \pm  0.000005 $ & -29 00 29.8902 $\pm 00.0001 $ & $ 1700.77 \pm    0.07 $ & $ -1679.00 \pm    0.10 $ \\ 
    \dots &  X & 17 45 40.163891 $ \pm  0.000014 $ & -29 00 29.8904 $\pm 00.0002 $ & $ 1700.62 \pm    0.18 $ & $ -1679.20 \pm    0.20 $ \\ 
    56750 &  K & 17 45 40.163889 $ \pm  0.000004 $ & -29 00 29.8903 $\pm 00.0001 $ & $ 1700.94 \pm    0.05 $ & $ -1678.49 \pm    0.10 $ \\ 
    \dots &  Q & 17 45 40.163892 $ \pm  0.000003 $ & -29 00 29.8901 $\pm 00.0001 $ & $ 1700.98 \pm    0.04 $ & $ -1678.29 \pm    0.10 $ \\ 
    56772 &  X & 17 45 40.163900 $ \pm  0.000016 $ & -29 00 29.8900 $\pm 00.0002 $ & $ 1701.27 \pm    0.21 $ & $ -1677.86 \pm    0.20 $ \\ 
    \dots & Ku & 17 45 40.163858 $ \pm  0.000008 $ & -29 00 29.8903 $\pm 00.0001 $ & $ 1700.72 \pm    0.10 $ & $ -1678.16 \pm    0.10 $ \\ 
    56892 & Ku & 17 45 40.163850 $ \pm  0.000028 $ & -29 00 29.8902 $\pm 00.0003 $ & $ 1701.65 \pm    0.37 $ & $ -1676.15 \pm    0.30 $ \\ 
    56899 &  K & 17 45 40.163904 $ \pm  0.000025 $ & -29 00 29.8902 $\pm 00.0003 $ & $ 1702.42 \pm    0.33 $ & $ -1676.06 \pm    0.30 $ \\ 
    \dots &  Q & 17 45 40.163868 $ \pm  0.000005 $ & -29 00 29.8900 $\pm 00.0001 $ & $ 1701.95 \pm    0.07 $ & $ -1675.90 \pm    0.10 $ \\ 
    \hline
    57073 &  X & 17 45 40.163893 $ \pm  0.000047 $ & -29 00 29.8905 $\pm 00.0004 $ & $ 1703.78 \pm    0.62 $ & $ -1673.72 \pm    0.40 $ \\ 
    \dots & Ku & 17 45 40.163846 $ \pm  0.000010 $ & -29 00 29.8898 $\pm 00.0002 $ & $ 1703.16 \pm    0.13 $ & $ -1673.02 \pm    0.20 $ \\ 
    57220 & Ku & 17 45 40.163862 $ \pm  0.000016 $ & -29 00 29.8894 $\pm 00.0003 $ & $ 1704.65 \pm    0.21 $ & $ -1670.33 \pm    0.30 $ \\ 
    \dots &  K & 17 45 40.163824 $ \pm  0.000026 $ & -29 00 29.8893 $\pm 00.0006 $ & $ 1704.15 \pm    0.34 $ & $ -1670.28 \pm    0.60 $ \\ 
    57387 & Ku & 17 45 40.163806 $ \pm  0.000004 $ & -29 00 29.8895 $\pm 00.0001 $ & $ 1705.36 \pm    0.05 $ & $ -1667.87 \pm    0.10 $ \\ 
    \dots &  K & 17 45 40.163800 $ \pm  0.000003 $ & -29 00 29.8893 $\pm 00.0001 $ & $ 1705.28 \pm    0.04 $ & $ -1667.72 \pm    0.10 $ \\ 
    57564 & Ku & 17 45 40.163708 $ \pm  0.000004 $ & -29 00 29.8893 $\pm 00.0001 $ & $ 1705.60 \pm    0.05 $ & $ -1664.98 \pm    0.10 $ \\ 
    \dots &  K & 17 45 40.163696 $ \pm  0.000004 $ & -29 00 29.8891 $\pm 00.0001 $ & $ 1705.44 \pm    0.05 $ & $ -1664.84 \pm    0.10 $ \\ 
    \dots &  X & 17 45 40.163781 $ \pm  0.000015 $ & -29 00 29.8897 $\pm 00.0003 $ & $ 1704.77 \pm    0.20 $ & $ -1668.60 \pm    0.30 $ \\ 
    57718 & Ku & 17 45 40.163653 $ \pm  0.000030 $ & -29 00 29.8886 $\pm 00.0005 $ & $ 1706.21 \pm    0.39 $ & $ -1661.98 \pm    0.50 $ \\ 
    \dots &  K & 17 45 40.163619 $ \pm  0.000015 $ & -29 00 29.8883 $\pm 00.0003 $ & $ 1705.76 \pm    0.20 $ & $ -1661.68 \pm    0.30 $ \\ 
    57923 & Ku & 17 45 40.163609 $ \pm  0.000006 $ & -29 00 29.8887 $\pm 00.0001 $ & $ 1707.41 \pm    0.08 $ & $ -1658.95 \pm    0.09 $ \\ 
    58060 & Ku & 17 45 40.163611 $ \pm  0.000006 $ & -29 00 29.8883 $\pm 00.0001 $ & $ 1708.62 \pm    0.08 $ & $ -1656.45 \pm    0.08 $ \\ 
    \dots &  K & 17 45 40.163574 $ \pm  0.000005 $ & -29 00 29.8886 $\pm 00.0001 $ & $ 1708.14 \pm    0.07 $ & $ -1656.76 \pm    0.09 $ \\ 
    58062 & Ku & 17 45 40.163561 $ \pm  0.000010 $ & -29 00 29.8884 $\pm 00.0001 $ & $ 1707.99 \pm    0.13 $ & $ -1656.45 \pm    0.12 $ \\ 
    \dots &  K & 17 45 40.163566 $ \pm  0.000008 $ & -29 00 29.8884 $\pm 00.0002 $ & $ 1708.05 \pm    0.10 $ & $ -1656.51 \pm    0.15 $ \\ 
    58070 & Ku & 17 45 40.163554 $ \pm  0.000011 $ & -29 00 29.8887 $\pm 00.0002 $ & $ 1707.96 \pm    0.14 $ & $ -1656.63 \pm    0.15 $ \\ 
    \dots &  K & 17 45 40.163561 $ \pm  0.000006 $ & -29 00 29.8885 $\pm 00.0001 $ & $ 1708.06 \pm    0.08 $ & $ -1656.42 \pm    0.11 $ \\ 
    58137 &  K & 17 45 40.163532 $ \pm  0.000026 $ & -29 00 29.8883 $\pm 00.0006 $ & $ 1708.25 \pm    0.34 $ & $ -1655.21 \pm    0.60 $ \\ 
    58178 &  K & 17 45 40.163528 $ \pm  0.000008 $ & -29 00 29.8881 $\pm 00.0002 $ & $ 1708.55 \pm    0.10 $ & $ -1654.46 \pm    0.16 $ \\ 
    58274 &  K & 17 45 40.163514 $ \pm  0.000010 $ & -29 00 29.8885 $\pm 00.0003 $ & $ 1709.20 \pm    0.13 $ & $ -1653.36 \pm    0.29 $ \\ 
    58318 &  K & 17 45 40.163528 $ \pm  0.000015 $ & -29 00 29.8881 $\pm 00.0003 $ & $ 1709.76 \pm    0.20 $ & $ -1652.25 \pm    0.33 $ \\ 
    58405 & Ku & 17 45 40.163542 $ \pm  0.000015 $ & -29 00 29.8871 $\pm 00.0002 $ & $ 1710.70 \pm    0.20 $ & $ -1649.99 \pm    0.20 $ \\ 
    \dots &  K & 17 45 40.163482 $ \pm  0.000014 $ & -29 00 29.8877 $\pm 00.0003 $ & $ 1709.91 \pm    0.18 $ & $ -1650.54 \pm    0.30 $ \\ 
    58406 & Ku & 17 45 40.163542 $ \pm  0.000015 $ & -29 00 29.8871 $\pm 00.0002 $ & $ 1710.71 \pm    0.20 $ & $ -1649.98 \pm    0.20 $ \\ 
    \dots &  K & 17 45 40.163492 $ \pm  0.000014 $ & -29 00 29.8880 $\pm 00.0003 $ & $ 1710.05 \pm    0.18 $ & $ -1650.88 \pm    0.25 $ \\ 

\enddata
\tablecomments{Horizontal line indicates the separation between previously published \citep{Bower15,Bower15erratum} and new data.}
\end{deluxetable}

\begin{deluxetable}{rrrr}
\tabletypesize{\footnotesize}
\tablecaption{Observed Sizes of \psr\ at 2cm wavelength \label{tab:sizes}}
\tablehead{
\colhead{MJD} & \colhead{Major Axis} & \colhead{Minor Axis} & \colhead{Position Angle} 
\\
              & \colhead{(mas)} &   \colhead{(mas)} & \colhead{(deg)} 
}
\startdata
56473 &  $ 5.4^{+0.5}_{-0.5} $ & $ 2.3^{+1.0}_{-2.0} $ & $  78^{+ 10}_{-  9} $ \\ 
56658 &  $ 4.9^{+0.6}_{-0.5} $ & $ 3.0^{+1.1}_{-1.7} $ & $  78^{+ 17}_{- 16} $ \\ 
56710 &  $ 5.0^{+0.3}_{-0.3} $ & $ 2.1^{+0.7}_{-1.1} $ & $  84^{+  6}_{-  7} $ \\ 
56772 &  $ 5.5^{+0.5}_{-0.5} $ & $ 3.6^{+0.8}_{-1.0} $ & $  79^{+ 13}_{- 13} $ \\ 
56892 &  $ 7.0^{+1.9}_{-2.1} $ & $ 2.3^{+3.1}_{-2.3} $ & $  77^{+ 28}_{- 24} $ \\ 
\hline
57387 &  $ 5.0^{+0.3}_{-0.3} $ & $ 2.9^{+0.6}_{-0.8} $ & $  80^{+  9}_{-  8} $ \\ 
57564 &  $ 5.3^{+0.3}_{-0.3} $ & $ 2.6^{+0.5}_{-0.6} $ & $  83^{+  6}_{-  5} $ \\ 
57923 &  $ 5.1^{+0.5}_{-0.5} $ & $ 2.7^{+0.7}_{-0.9} $ & $  75^{+ 10}_{-  9} $ \\ 
58070 &  $ 4.7^{+0.5}_{-4.7} $ & $ 1.6^{+2.9}_{-1.6} $ & $  85^{+  9}_{-  9} $ \\ 
\hline
\hline
Average & $ 5.2 \pm 0.2 $ & $ 2.7 \pm 0.4   $ & $  81 \pm   3 $ \\ 

\enddata
\tablecomments{Horizontal line indicates the separation between previously published \citep{Bower15,Bower15erratum} and new data.}
\end{deluxetable}

\section{Astrometric Analysis}

We analyze the relative astrometric data in the same manner as in \citet{Bower15}.  Fitting is done relative to the following functional form:
\begin{eqnarray}
\Delta\alpha & = & \Delta\alpha_0 + f_\alpha(\pi,\alpha,\delta) + \mu_\alpha * ({\rm MJD - MJD_0}) + {1 \over 2} a_\alpha * ({\rm MJD - MJD_0})^2 - \Phi_\alpha \lambda \\
\Delta\delta & = & \Delta\delta_0 + f_\delta(\pi,\alpha,\delta) + \mu_\delta * ({\rm MJD - MJD_0}) + {1 \over 2} a_\delta * ({\rm MJD - MJD_0})^2 - \Phi_\delta \lambda \nonumber,
\label{eqn:astrofit}
\end{eqnarray}

\begin{figure}[htb]
\includegraphics[width=0.75\textwidth]{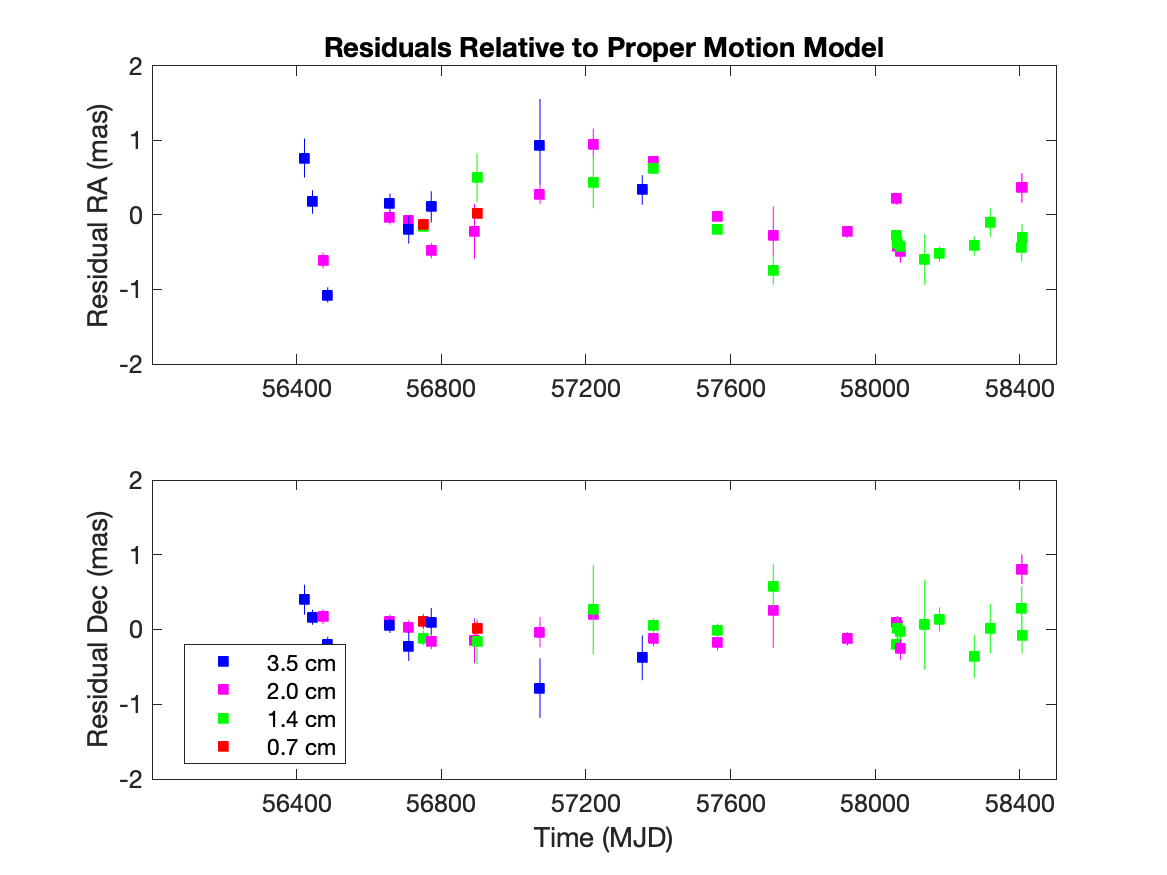}
\includegraphics[width=0.75\textwidth]{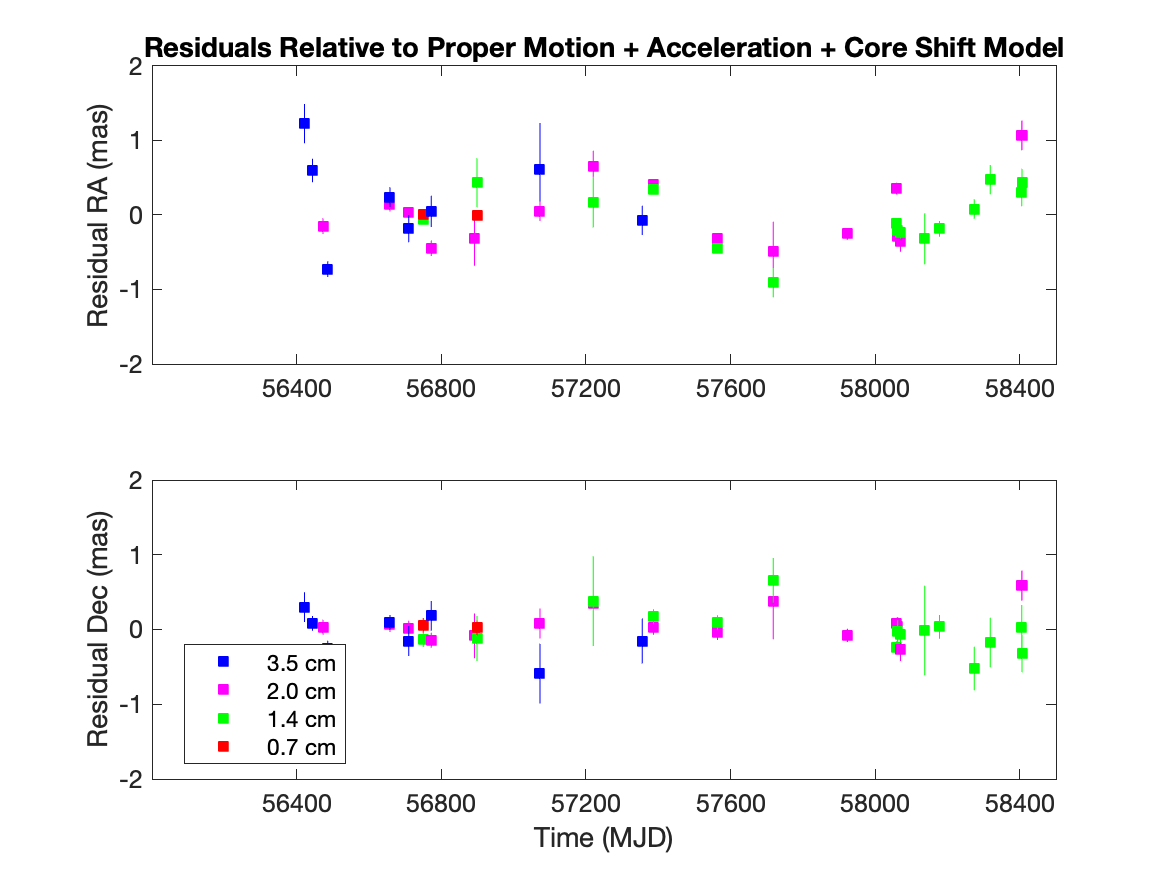}
\caption{Astrometric residuals after the fit for proper motion (upper left), 
and after the fit for proper motion, acceleration, and core shift (lower left).
\label{fig:residual}
}
\end{figure}

Fitted terms include the parallax ($\pi$), proper  motion ($\mu_\alpha$, $\mu_\delta$), acceleration ($a_\alpha$, $a_\delta$), and core shift ($\Phi_\alpha$, $\Phi_\delta$) as a function of wavelength $\lambda$.  $(f_\alpha,f_\delta)$ are the predictions for the effect of parallax $\pi$ at a given position $(\alpha,\delta)$, relative to the parallax of \sgra.  The mean epoch of the data is ${\rm MJD_0}=57424.7$.  
Fits are obtained for progressively more complex sets of parameters:  first, proper motion only; second, proper motion plus acceleration; third, proper motion plus core shift;  fourth, proper motion plus acceleration and core shift; and fifth, proper motion plus parallax.  The results of the first four fits are summarized in Table~\ref{tab:fits} with residuals to principal models shown in Figure~\ref{fig:residual}.  The first two fits are performed with both bootstrap and least squares (LSQ) methods as a means of confirming the accuracy of error estimates (Figure~\ref{fig:pmbootstrap}). The bootstrap method uses $10^4$ astrometric series via sampling with replacement.  We find good consistency in estimates of parameters and their errors across different fits and methods.

\begin{deluxetable}{llrrrrrr}
\tablecaption{Astrometric Fits for \psr\ \label{tab:fits}}
\tablehead{
\colhead{Parameter} & \colhead{Units} & \colhead{PM Bootstrap} &
\colhead{PM Bootstrap + Accel.} & \colhead{PM LSQ} & \colhead{PM + Accel.}
& \colhead{PM + Core Shift}
& \colhead{PM + Accel + Core Shift}
}
\startdata
$\Delta\alpha_0$ & (mas)    & $ 1704.86 \pm   0.11 $ & $ 1705.12 \pm   0.17 $ & $ 1704.86 \pm   0.07 $ & $ 1705.15 \pm   0.08 $ &$ 1704.89 \pm   0.18 $ &$ 1705.09 \pm   0.15 $  \\ 
$\Delta\delta_0$ & (mas)    & $ -1667.13 \pm   0.03 $ & $ -1667.27 \pm   0.05 $ & $ -1667.13 \pm   0.03 $ & $ -1667.27 \pm   0.05 $ &$ -1667.18 \pm   0.09 $ &$ -1667.24 \pm   0.08 $  \\ 
$\mu_\alpha$ & (mas/yr)      & $ 2.04 \pm 0.04 $ & $ 2.01 \pm 0.03 $ & $ 2.04 \pm 0.04 $ & $ 2.01  \pm 0.04 $ & $ 2.04  \pm 0.04 $ & $ 2.01  \pm 0.04 $ \\ 
$\mu_\delta$ & (mas/yr)      & $ 6.07 \pm 0.02 $ & $ 6.09 \pm 0.02 $ & $ 6.07 \pm 0.02 $ & $ 6.09  \pm 0.02 $ & $ 6.08  \pm 0.02 $ & $ 6.09  \pm 0.02 $ \\ 
$a_\alpha$ & (mas/yr$^2$)      & \dots & $ -0.238  \pm 0.089 $ & \dots & $ -0.252  \pm 0.052 $ & \dots & $ -0.256  \pm 0.054 $ \\ 
$a_\delta$ & (mas/yr$^2$)      & \dots & $ 0.089  \pm 0.033 $ & \dots & $ 0.089  \pm 0.027 $ & \dots & $ 0.094  \pm 0.029 $ \\ 
$\Phi_\alpha$ & (mas/cm)      & \dots & \dots & \dots & \dots & $ 0.02  \pm 0.10 $& $ -0.04  \pm 0.08 $ \\ 
$\Phi_\delta$ & (mas/cm)      & \dots & \dots & \dots & \dots & $ -0.03  \pm 0.04 $& $ 0.02  \pm 0.04 $ \\ 
$ \chi^2_\alpha /d.o.f._\alpha$ &  & \dots & \dots & $ 852.9 / 39 $ &$ 530.1 / 38 $ &$ 851.9 / 38 $ &$ 527.3 / 37 $ \\ 
$ \chi^2_\delta /d.o.f._\delta$ &  & \dots & \dots & $ 85.4 / 39 $ &$ 66.0 / 38 $ &$ 84.2 / 38 $ &$ 65.7 / 37 $ \\ 

\hline
\multicolumn{8}{c}{Fits without first four epochs} \\
\hline
$\Delta\alpha_0$ & (mas)    & $ 1704.88 \pm   0.11 $ & $ 1705.11 \pm   0.18 $ & $ 1704.88 \pm   0.06 $ & $ 1705.14 \pm   0.08 $ &$ 1704.71 \pm   0.19 $ &$ 1705.01 \pm   0.17 $  \\ 
$\Delta\delta_0$ & (mas)    & $ -1667.14 \pm   0.03 $ & $ -1667.27 \pm   0.06 $ & $ -1667.14 \pm   0.03 $ & $ -1667.27 \pm   0.05 $ &$ -1667.17 \pm   0.09 $ &$ -1667.27 \pm   0.09 $  \\ 
$\mu_\alpha$ & (mas/yr)      & $ 2.00 \pm 0.04 $ & $ 2.00 \pm 0.03 $ & $ 2.01 \pm 0.04 $ & $ 2.00  \pm 0.04 $ & $ 2.00  \pm 0.04 $ & $ 2.00  \pm 0.04 $ \\ 
$\mu_\delta$ & (mas/yr)      & $ 6.09 \pm 0.02 $ & $ 6.09 \pm 0.02 $ & $ 6.09 \pm 0.02 $ & $ 6.09  \pm 0.02 $ & $ 6.09  \pm 0.02 $ & $ 6.09  \pm 0.02 $ \\ 
$a_\alpha$ & (mas/yr$^2$)      & \dots & $ -0.223  \pm 0.103 $ & \dots & $ -0.242  \pm 0.057 $ & \dots & $ -0.237  \pm 0.057 $ \\ 
$a_\delta$ & (mas/yr$^2$)      & \dots & $ 0.094  \pm 0.042 $ & \dots & $ 0.096  \pm 0.033 $ & \dots & $ 0.096  \pm 0.034 $ \\ 
$\Phi_\alpha$ & (mas/cm)      & \dots & \dots & \dots & \dots & $ -0.11  \pm 0.11 $& $ -0.08  \pm 0.10 $ \\ 
$\Phi_\delta$ & (mas/cm)      & \dots & \dots & \dots & \dots & $ -0.02  \pm 0.05 $& $ 0.00  \pm 0.05 $ \\ 
$ \chi^2_\alpha /d.o.f._\alpha$ &  & \dots & \dots & $ 681.6 / 35 $ &$ 445.7 / 34 $ &$ 662.5 / 34 $ &$ 436.6 / 33 $ \\ 
$ \chi^2_\delta /d.o.f._\delta$ &  & \dots & \dots & $ 69.8 / 35 $ &$ 56.0 / 34 $ &$ 69.5 / 34 $ &$ 56.0 / 33 $ \\ 

\hline
\multicolumn{8}{c}{Fits with estimated systematic uncertainty included} \\
\hline
$\Delta\alpha_0$ & (mas)    & $ 1704.78 \pm   0.07 $ & $ 1704.96 \pm   0.13 $ & $ 1704.78 \pm   0.07 $ & $ 1704.96 \pm   0.12 $ &$ 1704.69 \pm   0.20 $ &$ 1704.80 \pm   0.20 $  \\ 
$\Delta\delta_0$ & (mas)    & $ -1667.11 \pm   0.04 $ & $ -1667.27 \pm   0.06 $ & $ -1667.11 \pm   0.04 $ & $ -1667.27 \pm   0.06 $ &$ -1667.17 \pm   0.11 $ &$ -1667.24 \pm   0.10 $  \\ 
$\mu_\alpha$ & (mas/yr)      & $ 2.01 \pm 0.04 $ & $ 2.01 \pm 0.04 $ & $ 2.01 \pm 0.04 $ & $ 2.01  \pm 0.04 $ & $ 2.02  \pm 0.04 $ & $ 2.03  \pm 0.04 $ \\ 
$\mu_\delta$ & (mas/yr)      & $ 6.09 \pm 0.02 $ & $ 6.10 \pm 0.02 $ & $ 6.09 \pm 0.02 $ & $ 6.10  \pm 0.02 $ & $ 6.09  \pm 0.02 $ & $ 6.09  \pm 0.02 $ \\ 
$a_\alpha$ & (mas/yr$^2$)      & \dots & $ -0.113  \pm 0.074 $ & \dots & $ -0.111  \pm 0.057 $ & \dots & $ -0.125  \pm 0.059 $ \\ 
$a_\delta$ & (mas/yr$^2$)      & \dots & $ 0.098  \pm 0.037 $ & \dots & $ 0.100  \pm 0.030 $ & \dots & $ 0.104  \pm 0.032 $ \\ 
$\Phi_\alpha$ & (mas/cm)      & \dots & \dots & \dots & \dots & $ -0.04  \pm 0.10 $& $ -0.09  \pm 0.10 $ \\ 
$\Phi_\delta$ & (mas/cm)      & \dots & \dots & \dots & \dots & $ -0.03  \pm 0.05 $& $ 0.02  \pm 0.05 $ \\ 
$ \chi^2_\alpha /d.o.f._\alpha$ &  & \dots & \dots & $ 51.1 / 39 $ &$ 46.4 / 38 $ &$ 50.9 / 38 $ &$ 45.3 / 37 $ \\ 
$ \chi^2_\delta /d.o.f._\delta$ &  & \dots & \dots & $ 45.3 / 39 $ &$ 34.9 / 38 $ &$ 44.9 / 38 $ &$ 34.8 / 37 $ \\ 

\enddata
\end{deluxetable}

The LSQ methods produce a $\chi_\nu^2$ statistic that can be used to validate the quality of the fit.   Declination fits improve with the introduction of an acceleration term but do not improve with introduction of the core shift term.  Residuals remain significantly in excess of 1, however, indicative of the presence of systematic errors or underestimated thermal errors.  When we add in quadrature a systematic uncertainty of $\sim 0.15$ mas to all measurements, we find $\chi^2_\nu\sim 1$ for all declination fits.

Right ascension fits, on the other hand, show $\chi_\nu^2 >>1$. In particular, and as already noted by \citet{Bower15}, the first four epochs appear substantially discrepant.  We test the contribution of the first four epochs to the final results by excluding them from 
the fitting.  These results are shown in the middle section of Table~\ref{tab:fits}.  While exclusion of these first four epochs does improve $\chi^2_\nu$, we find that there still is a significant contribution from the full range of epochs.  We find that inclusion of a right ascension systematic uncertainty term of 0.35 mas to each epoch produces $\chi^2_\nu\sim 1$.

The $\sim$ 2:1 ratio of these systematic error estimates suggests an origin associated with the scattering medium, which produces images that are elongated East-West with a 2:1 axial ratio, unlike the synthesized beam, which is elongated North-South.  The scattering size is $\sim 1.4$ mas in the major axis at a wavelength of 1.3 cm.  Accordingly, for the substantial majority of these measurements, this systematic error term is consistent with a scaling of 1/4 of the beam, or less.  While many of the plausible sources of error could reasonably be expected to scale positively with the observing wavelength (as the apparent source size is also increasing with wavelength), we note that we see no evidence that this is the case. However, the observations with both the lowest and highest observing wavelengths are clustered in time early in the observing campaign, meaning that separating time dependent effects from wavelength dependent effects is not straightforward. \added{Selection of only the Ku band data, which is often of the highest significance, does not reduce the scatter in right ascension residuals.}

The choice of a reduced data set or the inclusion of systematic error changes the precise values of fitted parameters but does not change these values systematically or significantly.  Results for proper motion, acceleration, and core shift in both coordinates remain within $1\sigma$ of each other regardless of the choices that are made.

\section{Results}

\subsection{Astrometric Constraints}

The proper motion constraint obtained through these measurements is  independent within the errors of whether we fit for additional terms for acceleration, parallax, or core shift.  The proper motion value is also consistent with the measurement previously reported in \citet{Bower15} with a five-fold reduction in the measurement uncertainty.  Thus, the central conclusion remains that the proper motion of the magnetar is consistent with neighboring stars in the clockwise (CW) disk of stars and supports the conclusion that the magnetar originated from this stellar disk with a modest initial velocity kick.

\begin{figure}[htb]
\includegraphics[width=0.5\textwidth]{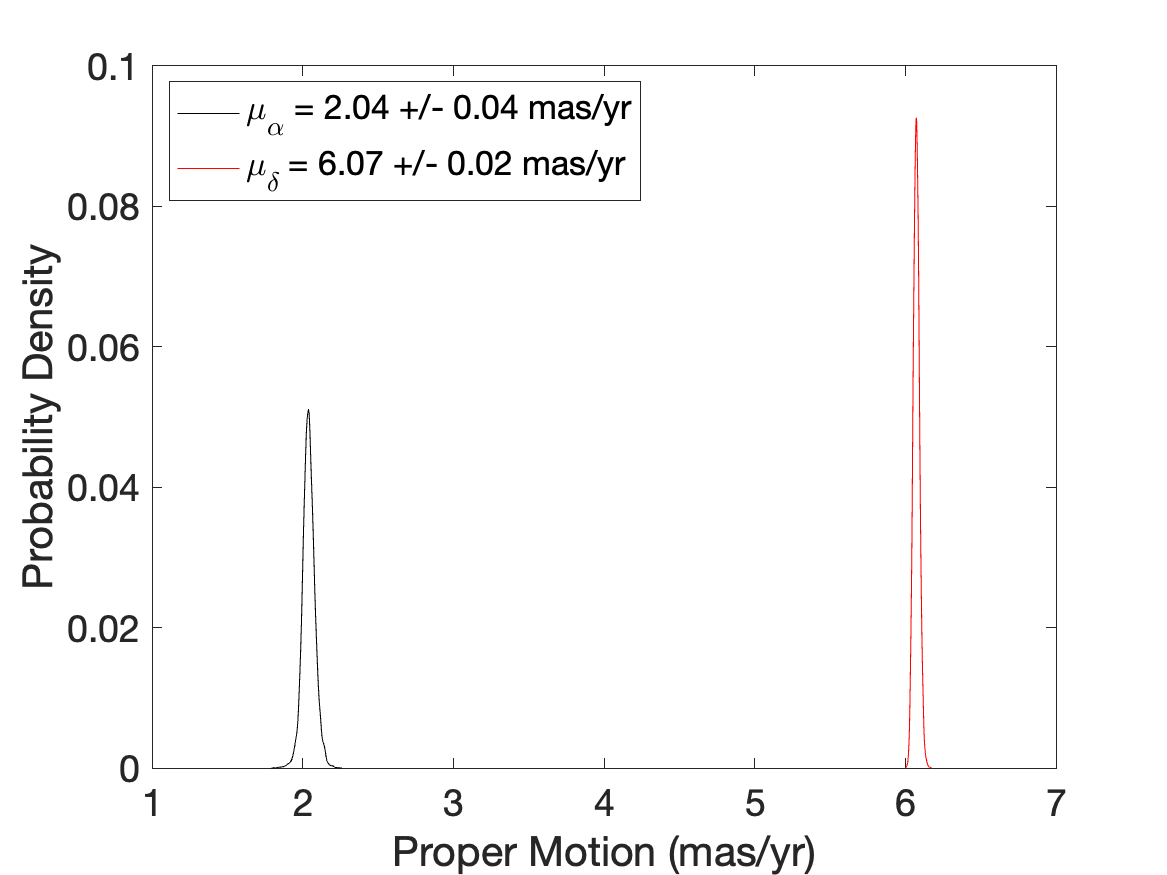}
\includegraphics[width=0.5\textwidth]{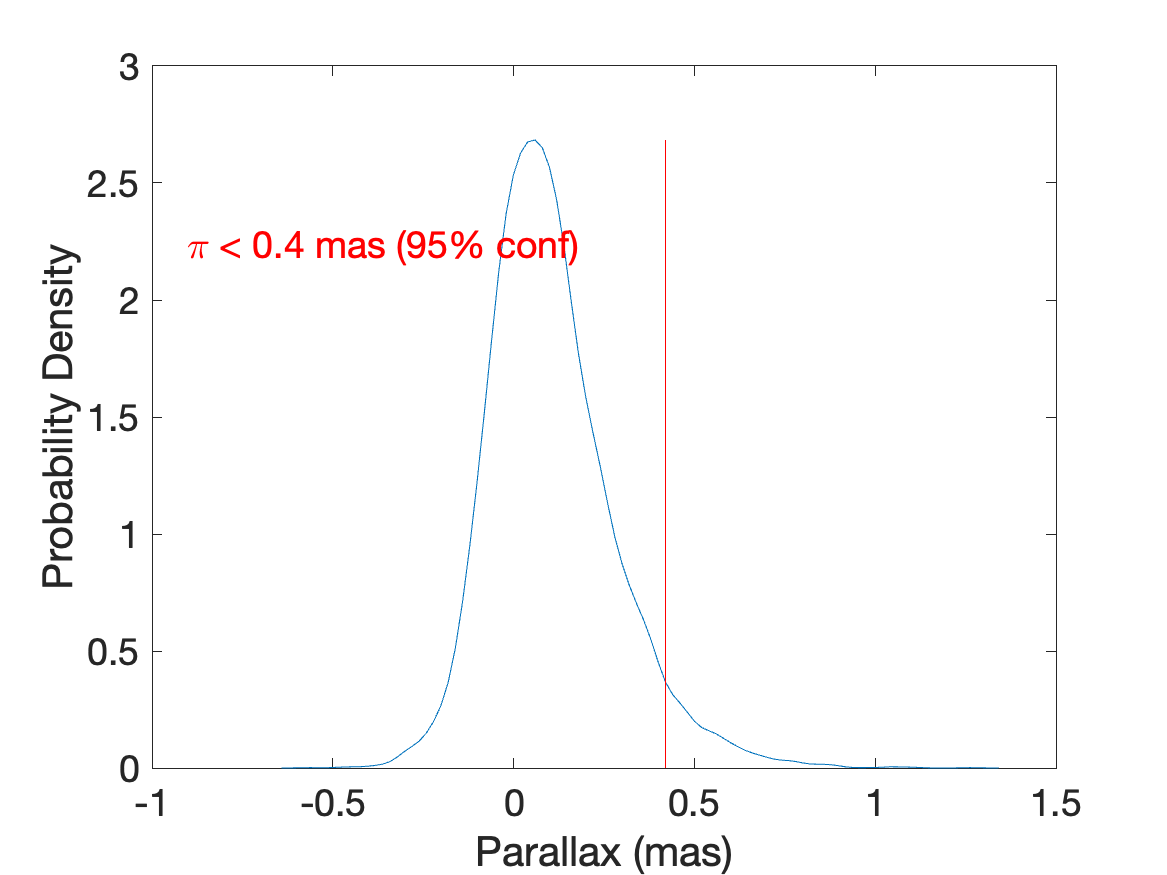}
\includegraphics[width=0.5\textwidth]{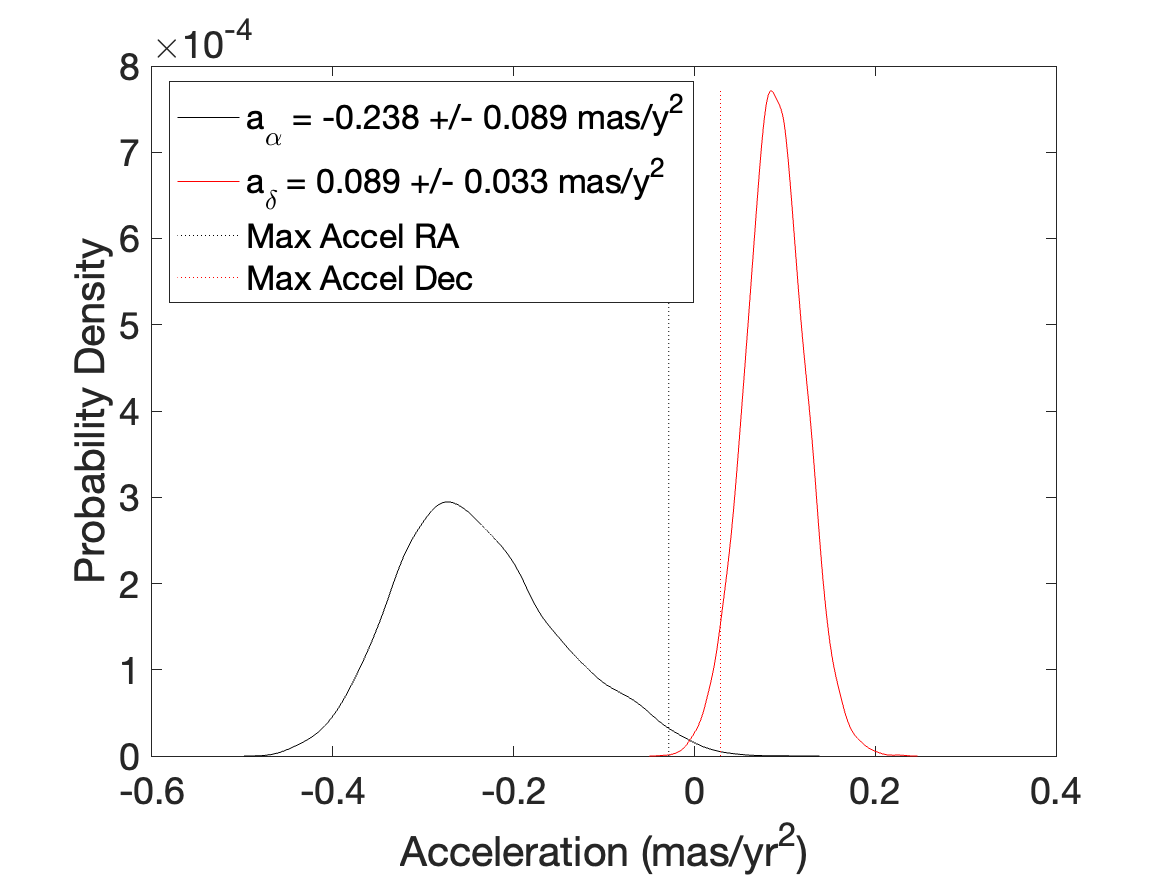}
\caption{Empirical probability density distribution of proper motion solutions from the bootstrap resampling method (upper left).
Distribution of the measured parallax $\pi$ of the magnetar referenced to \sgra\ (upper right).  The upper limit at 95\% confidence of $\pi<0.6$ mas is consistent with the expectation that the magnetar is located in the GC.
Empirical probability density distribution of acceleration
from the bootstrap resampling method (lower left).  Vertical dotted lines indicate the maximum acceleration terms as the result of acceleration by \sgra.  
\label{fig:pmbootstrap}
}
\end{figure}

For the expected case of the pulsar in orbit around \sgra at a radius $r \lsim 1$ pc, we expect the measured parallax relative to \sgra\ to be \added{reduced by a factor of $r/D\lsim 10^{-5} \times$ the expected value for a distance background source, where $D$ is the distance from the Sun to the Galactic Center, leading to an expected value on the scale of nanoarcseconds}.  The current constraint $\pi<0.4$ mas (Figure~\ref{fig:pmbootstrap}) is therefore consistent with the pulsar being located at the GC.

In the case that the magnetar is at a minimal separation from \sgra\ given by the projected distance, the maximal acceleration in each coordinate is \added{determined by the inverse-square law to be} $|a|\approx 0.029\,{\rm mas\,y^{-2}}$.  The uncertainty in the acceleration estimate exceeds this value, albeit only minimally in the case of the declination coordinate.  One can estimate the requirements for a future detection of the acceleration in the event that the magnetar re-brightens.  A second measurement of the proper motion with the obtained uncertainty of $(0.06, 0.02)$ mas/y in the two coordinates obtained at a separation of 10 years would provide a $(5,10)\sigma$ detection of the acceleration.  Such a detection would provide definitive confirmation that the magnetar is bound to \sgra\ and a direct measure of the separation of these two sources.  A sample of stars at similar radii have measured accelerations from over two decades of NIR astrometric monitoring 
\citep{2023A&A...670A..36Y}.

As noted above, the measured positions in declination are consistent with no variations from linear motion with the inclusion of a small systematic error.  On the other hand, 
residuals in right ascension show large variations that have the appearance of systematic changes.  In \citet{Bower15} we identify two sources of calibration error that could introduce systematic astrometric errors: first, errors in the scattering model for \sgra\ that are proportionally larger in right ascension than declination, and second, calibration of the phased VLA used in the first four epochs.  The latter of these does not apply to the long timescale variations seen in these data.  Astrophysical sources of systematic error could arise from changes in the apparent centroid position of \sgra\, 
which we discuss in Section\,\ref{sec:sgra}, refractive image wander, and the influence of a binary companion.

Refractive wander occurs as the result of changes of the large scale index of refraction due to the relative motion of the source and the turbulent scattering screen.  The refractive timescale $\tau_R\sim d\theta/v$, where $d$ is the distance from the Earth to the scattering screen, $\theta$ is the angular broadening, and $v$ is the relative velocity.  For $d\approx 3$ kpc and $v\approx 250\kms$, 
we find $\tau_R\gsim 300$ days at $\lambda=2$ cm.  Notably, refractive wander is wavelength dependent, whereas the systematic error in our residuals does not exhibit an obvious wavelength dependence.  This absence is reflected in the null detections of a systematic core shift, which scales with $\lambda$.  Further, if the plasma column probed by the magnetar sightline is broadly comparable to that probed by  \sgra\, the long-term stability of the apparent size \citep{Bower06,2022ApJ...926..108C} and proper motion \citep{2020ApJ...892...39R} of \sgra\ argues against substantial refractive wander effects.

\begin{figure}[htb]
\includegraphics{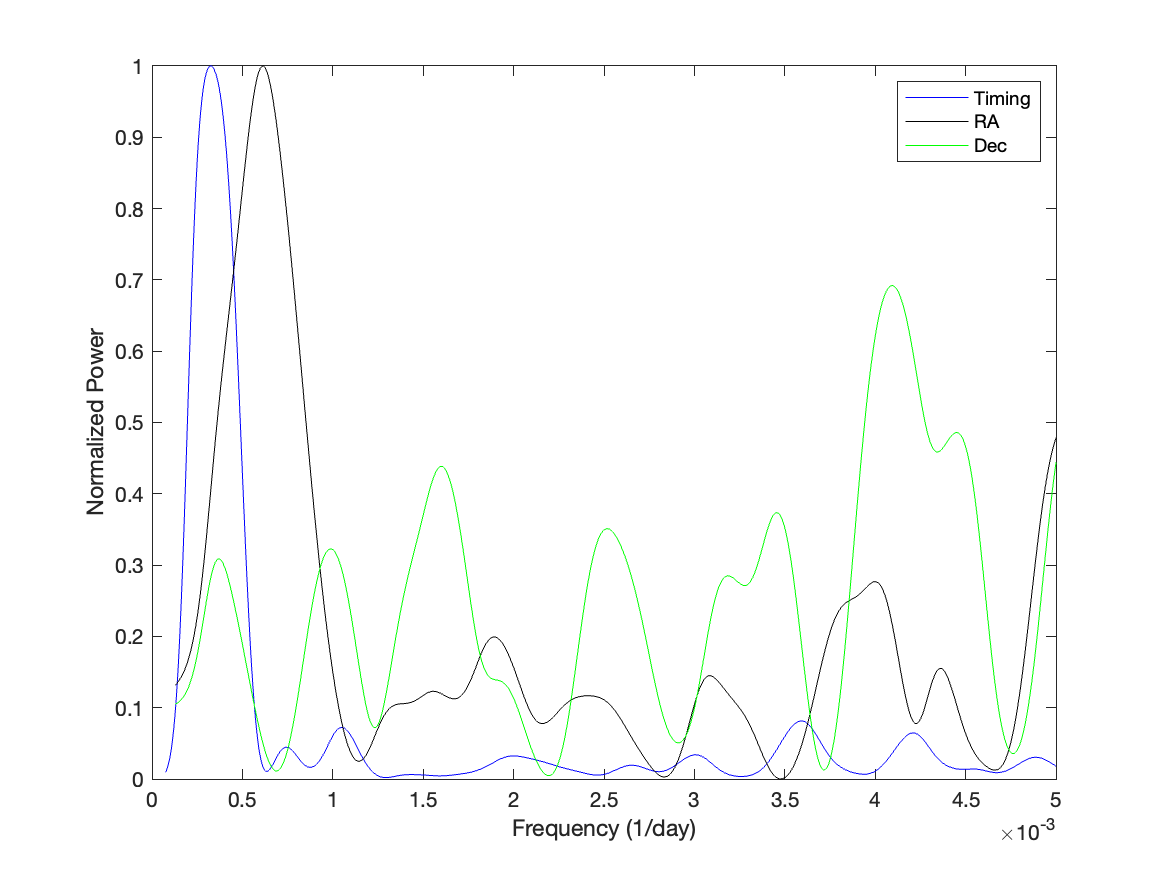}
\caption{Lomb-Scargle periodogram of the astrometric and timing data \citep{Eatoughetal}.  Periodic signals comparable to the data set length are often indicative of red noise variability.  A linear trend has been removed the timing residuals before calculation of the periodogram.  Calculation of the periodogram without the linear trend does not significantly alter the results.
\label{fig:lomb}
}
\end{figure}

\begin{figure}[htb]
\includegraphics{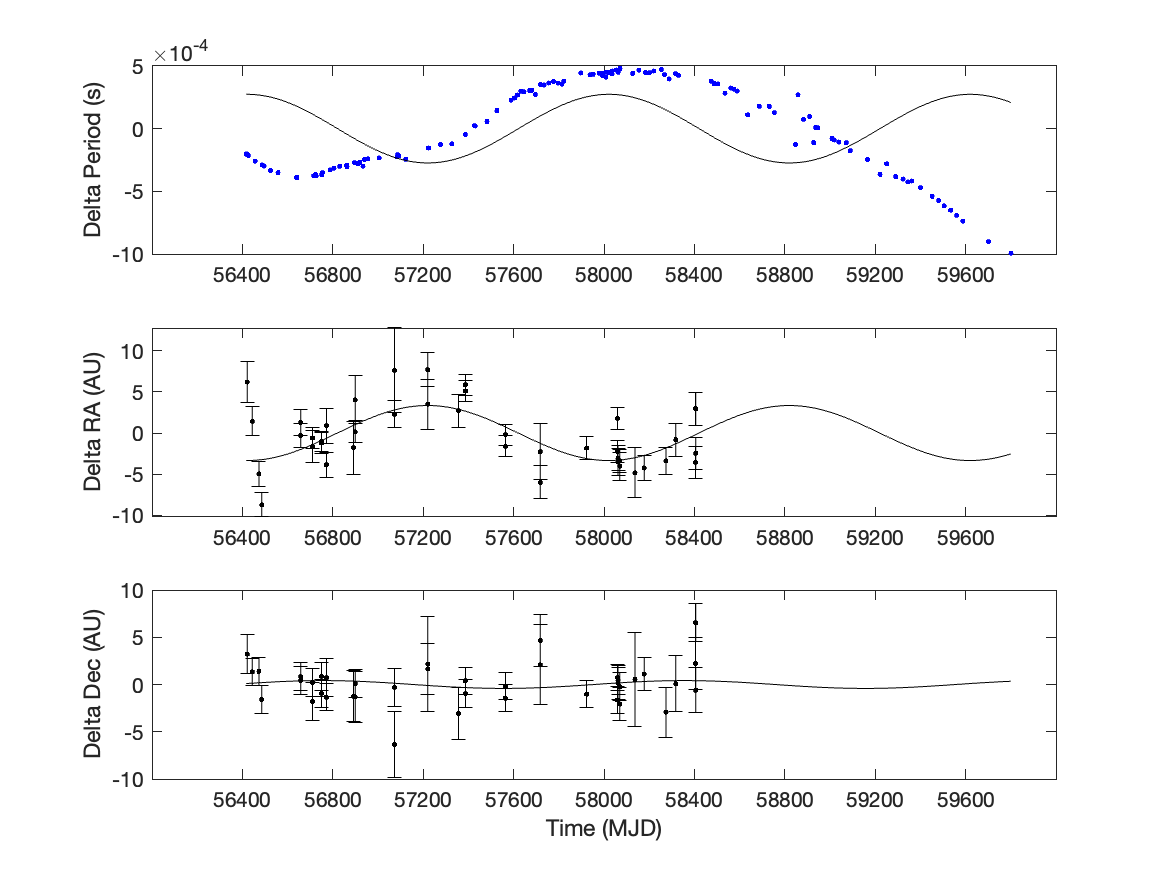}
\caption{The binary model (solid lines) from fitting astrometric data alone, plotted against the residual period from \citep{Eatoughetal} (top) and the astrometric data in right ascension (middle) and declination (bottom).  \added{The plotted binary model has a semimajor axis $a=3.3$ AU, inclination angle $i=-83^\circ$, and period $P=1600$ days.}  A linear trend has been removed from the period to produce the residual value.  While the magnitude of the variations are comparable between the period and astrometric data, there is no good correlation between the two.  More sophisticated modeling of variations in the period by \citet{Eatoughetal} limits the residuals to be much less than predicted by the VLBI orbital fit. 
\label{fig:astro+timing}
}
\end{figure}

We consider the possibility of a binary companion to the magnetar as the source of the astrometric residuals.  By eye, an apparent $\sim 1$-mas sinusoidal variation in the
right ascension can be seen, with a period comparable to the data set length.  A Lomb-Scargle periodogram analysis confirms that there is significant power on the longest timescale $\sim 1600$ days (Figure~\ref{fig:lomb}). 
For the declination residuals, we see no timescales on which there is significant periodicity up to the length of the data set.  
A  fit to the astrometric residuals for a binary companion reveals a set of orbits that match the broad features of the data (Figure~\ref{fig:astro+timing}). The best-fit solution finds a period $P=1600 \pm 200$ days, an inclination angle $i=95^\circ \pm 23^\circ$ (i.e., nearly face on), and a semi-major axis $a=3.3 \pm 1.0$ AU.  MCMC optimization shows broad distributions around these quantities.  The most highly inclined orbits provide the least constraint on the semi-major axis and eccentricity.  For a pulsar mass $M_P=1.4 \msun$, the companion mass has a median value $M_2=5.5 \msun$ but with $\pm 2\sigma$ values that range from $0.9 < M_2 < 35 \msun$.  High mass stellar companions are ruled out by NIR  imaging which have main-sequence detection limits of $\gsim 3 \msun$ \citep{2010RvMP...82.3121G,2020ApJ...896..100A}, meaning that any such putative companion would necessarily be a \added{neutron star or} black hole - possibly making \psr\ the first pulsar-black hole binary.

A binary companion will introduce a residual in timing of the pulsar signal with the same period as the astrometric variations.  A $10\msun$ companion at a separation of 1 mas would  accelerate the pulsar producing a change $\dot{P}\sim 10^{-11}$, which is comparable to the observed $\dot{P}=-6 \times 10^{12}$ \citep{2014ApJ...786...84K}.  
Timing measurements for the magnetar are difficult, however, given the unstable magnetar pulse profile, the possibility of glitches in the timing, the long-term reduction in the pulse flux density, and general timing noise seen in magnetars and young pulsars.
\added{Timing analysis} of an extensive data set spanning $\sim 3400$ days does not find evidence for a periodic signal consistent with the astrometric period \citep{Eatoughetal}.  
The periodogram of the pulse period reveals peak power at $\sim 3100$ days (Figure~\ref{fig:lomb}), nearly twice the period favoured by variations in the VLBI positions in right ascension.  In the case of both the astrometric and timing data, the peak period is close to the observational duration.  Additionally, the best-fit astrometric model is a poor fit to the timing data (Figure~\ref{fig:astro+timing}). Joint fits fail to find an adequate solution without the introduction of large systematic errors.
\citet{Eatoughetal} conduct 
a more extensive binary analysis of the timing data that incorporates red noise corresponding to intrinsic magnetar pulse properties and also find that a high mass companion to the magnetar is ruled out.

We conclude that no potential effect considered here provides a compelling explanation for the residual astrometric variations.  The inconsistency between the pulsar period measurements and the astrometry appears to eliminate a binary companion model.  The alternative explanations of refractive wander and variations in the morphology of \sgra\ could potentially produce residuals of the appropriate magnitude.  The absence of a clear signature of a frequency-dependent effect, however, undermines the case for refractive wander.   Structural changes in \sgra\ may also be inconsistent with the long-term accuracy of the independent astrometry and morphology measurements made of \sgra.  We do note, though, that many of those measurements are made at shorter wavelengths, where effects may be smaller.

\subsection{Angular Broadening and the Detectability of Pulsars in the GC}

\begin{figure}[htb]
\includegraphics{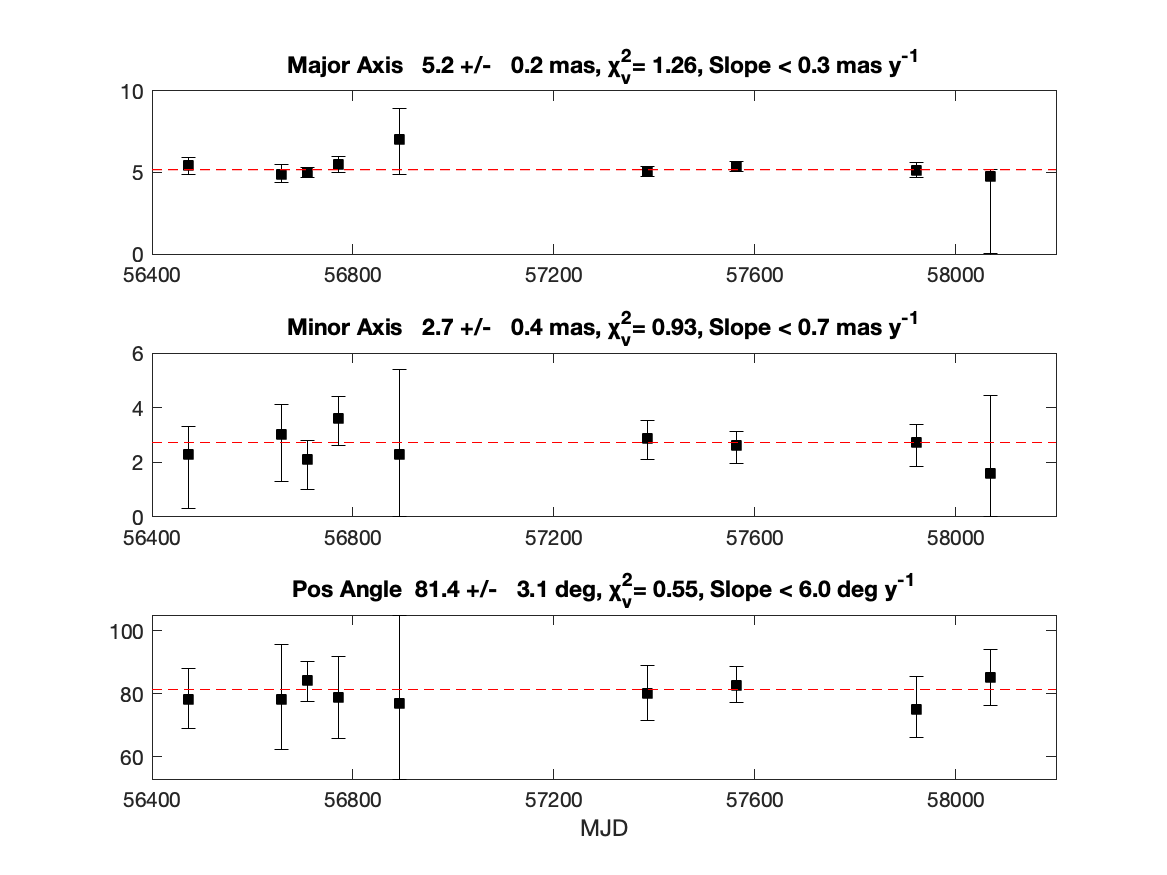}
\caption{
Apparent size of the magnetar at 2 cm wavelength as a function of time for Gaussian deconvolutions of the imaged size in major axis (top), minor axis (middle), and position angle (bottom) parameters.  Observations previously reported in \citet{Bower15} are before MJD 57000.  The red dashed line is the weighted mean for each parameter.
\label{fig:sizes}
}
\end{figure}

The apparent size \added{of the magnetar} as a function of time for the Ku band (15 GHz) is shown in Figure~\ref{fig:sizes}, including results previously reported.  For these high SNR epochs, we see no evidence for change in major axis size, minor axis size, and position angle over time.  The mean sizes are consistent with past measurements of the magnetar reported in \citet{Bower15} and, therefore, also with past measurements of \sgra.  
We set $3\sigma$ upper limits on a linear change in the major axis of $<0.3\,{\rm mas\, y^{-1}}$, approximately $<6\%$ per year.  

The sizes for the magnetar are derived following self-calibration of the visibilities on \sgra\ with a standard model. This self-calibration does not force the magnetar size to be the same as \sgra, but it {\em does} enforce a time-independent size for \sgra.  Thus, the results do rule out independent variations of the magnetar size relative to \sgra\ but they do not rule out coherent variations in the size of the two sources.  However, variations in the size of \sgra\ have been explored extensively without evidence for secular changes \citep{Bower06,2022ApJ...926..108C}.

The timescale for refractive variations at 15 GHz is $\gsim 300$ days, significantly shorter than the $>4$ y timescale of the magnetar measurements.  This stability is in contrast to the variations seen with other interstellar propagation measurements seen towards the magnetar.  Both the DM and the RM have been seen to vary over a range of timescales \citep{2018ApJ...852L..12D}.  
The DM has a component that arises from the  integrated electron population along the line of sight and varies by less than 1\% over 4 years.  On the other hand, the RM varies by $\sim 5\%$ on a range of timescales.  The majority of the RM likely arises from the central $\sim 10$ pc, consistent with the profile of electron densities traced by diffuse X-ray gas (and magnetic fields), with variations arising from the most turbulent components of the plasma at a distance of $\sim 0.1$ pc from \sgra\ \citep{2013Natur.501..391E}.  
The lack of similar variations in the scattering properties further supports the interpretation that the different propagation effects arise from different regions of the interstellar medium.  

The angular broadening is dominated by material arising at a distance of $\sim 5$ kpc from the GC, most likely associated with a supernova remnant in a spiral arm \citep{2014ApJ...780L...2B,2017MNRAS.471.3563D}.  Additionally, modeling the source structure as a function of pulse phase suggests additionally that the scattering is dominated by a single screen \citep{wucknitz2015probinginterstellarscatteringgalactic}.  This single scattering screen would be insufficient to obscure ordinary pulsars in the GC from existing searches.  Nevertheless, given the potential significance of scattering we consider the degree to which a second scattering screen can be constrained by these data.  We can take the upper limit on secular variation of the major axis (or the uncertainty in the major axis) as an estimate of the contribution of an additional, \added{time variable} GC component to the angular broadening.

The temporal scattering timescale $\tau$ for a \added{single} thin scattering screen is related to the angular broadening by the following equation:
\begin{equation}
\tau = 6.2 {\rm\ s\ } \times {\left( D \over 8.3 {\rm kpc} \right)} {\left( \theta_1 \over 1.3 {\rm\ arcsec} \right)^2} \left( {D \over \Delta} - 1 \right) \nu^{-4},
\label{eqn:tau}
\end{equation}
where $D = 8.3$ kpc \citep{2019Sci...365..664D,2022A&A...657L..12G}
is the distance to the GC, $\Delta$ is the distance from the pulsar to the scattering screen,
and $\nu$ is the observing frequency
in GHz \citep{1997ApJ...475..557C}.  $\theta_1$ is the observed angular size 
extrapolated to a frequency of 1.0 GHz using a scaling of $\nu^{-2}$.  If we substitute the mean scattering properties of the magnetar into Equation~\ref{eqn:tau}, we find a distance $\Delta=5.8 \pm 0.3$\, kpc.  

\added{Now, we can add in a second thin scattering screen closer to \sgra\ and at a sufficiently large separation from the dominant scattering screen at distance $\Delta$ such that the scattering angular diameters add in quadrature and the temporal broadening is the sum of the two properties of the two screens. We assume that this second scattering screen was not present at the beginning of the observations \psr\ when the scattering size was $\theta$ and was present at the end of observations.  Observationally, we constrain that the observed scattering size has not changed by more than the limit on the total secular change $\delta\theta$ in the source size. 
In this case, the maximum angular broadening attributed to the second scattering screen is $\theta_1^s=\sqrt{2\theta\delta\theta}=0.4$\,arcsec at 1 GHz.
We also require that the temporal broadening cannot have changed so much that the pulses would be temporally broadened significantly, which depends on the distance $\Delta^s$ of the second scattering screen from the GC.  Under these assumptions, a secondary scattering screen at $\Delta^s \lsim 3$ kpc would produce additional scattering $\tau^s > 1$\,sec, making the magnetar unobservable.  Smaller variations in the angular broadening, therefore, would be consistent with \psr\ remaining observable throughout these observations. An order of magnitude improvement in the uncertainty of the angular broadening would set a limit $\Delta^s \gsim 0.5$ kpc. }

Together these results find no evidence for the existence of a second strong scattering screen in the GC that would obscure a population of GC pulsars.  To accommodate such a model, the ionized ISM must be patchy on scales larger than \added{the distance traversed by the pulsar in the course of these observations} $\sim 200$ AU, larger than estimates of the inner scale of turbulence.  Based on diffractive properties of the scattering, the inner scale of turbulence is estimated to be $800 \pm 200$ km \citep{2018ApJ...865..104J}.  On larger scales, \citet{2013ApJ...773...67R} estimated patchiness in the ionized ISM of the GC to exist on a scale of 10\arcmin, approximately 20 pc.  Filaments and other features in the GC can have much smaller angular scales \citep{2022ApJ...925..165H}.

\subsection{Constraints on \sgra \label{sec:sgra}}

A wide variety of constraints have not yet resolved the question of whether \sgra\ has a jet feature.  Most recently, analysis of Event Horizon Telescope (EHT) images and polarimetric data with GRMHD models suggests that a small-scale jet may be present but definitive evidence does not exist \citep{2022ApJ...930L..16E,2024ApJ...964L..26E}.  One of the expected signatures of an inhomogeneous jet would be the presence of a core shift, i.e., a shift in the centroid of \sgra\ proportional to wavelength, as a result of variations in the optical depth \citep{1979ApJ...232...34B,2009A&A...496...77F,2014A&A...570A...7M,2023arXiv231212951F}.  This effect has been clearly seen in the jet of M87 \citep{2011Natur.477..185H}.  Since, the pulsar positions are measured relative to \sgra\ in each epoch, the wavelength-dependence of the positions can be used to constrain the core shift.

\begin{figure}[htb]
\includegraphics{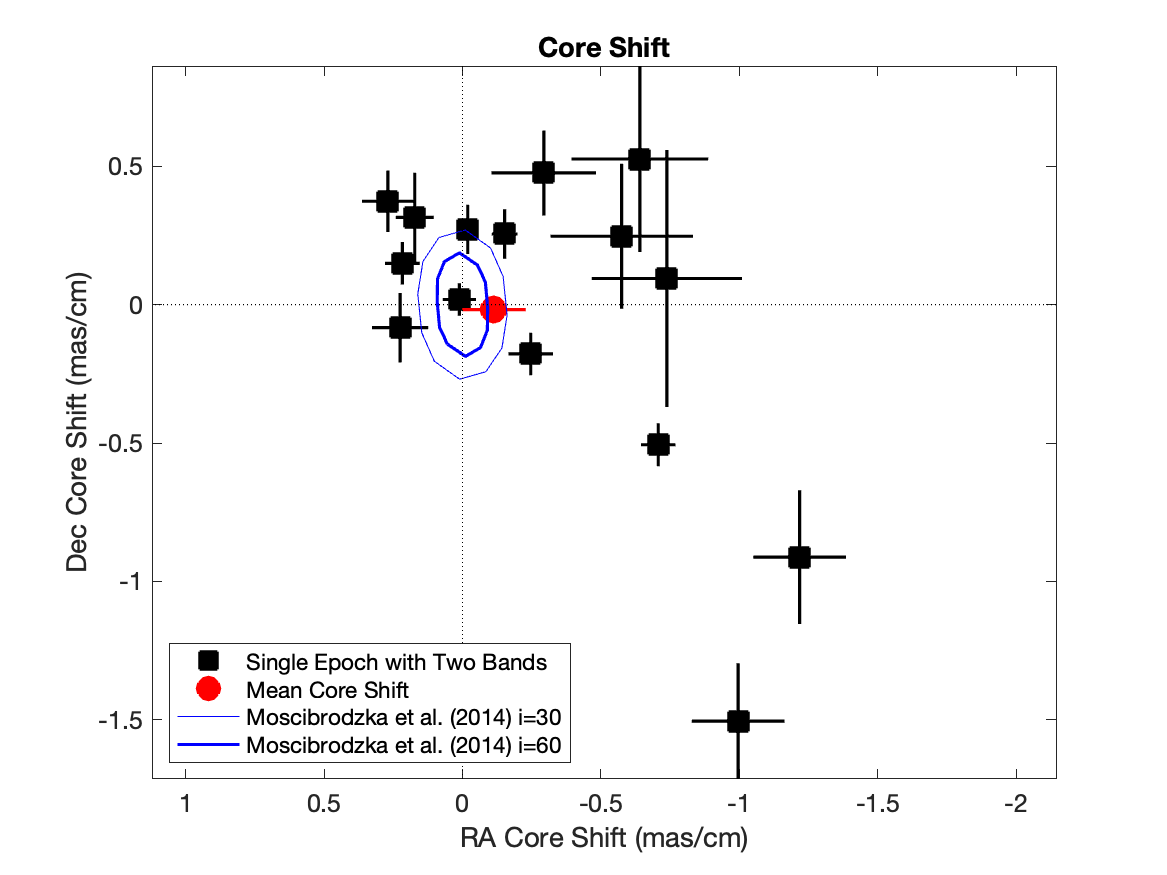}
\caption{Core shift for the  epochs with two frequency observations (black squares) and for the
fitted average over all epochs (red circle).  Note that the mean core shift includes data from epochs in which both one and two frequency band observations are included, but that single-epoch core shifts are estimated only when more than one frequency is present.  The light and dark blue curves show the core shift estimated from GRMHD jet simulations
with an inclination angle of $30^\circ$ and $60^\circ$ oriented towards different position angles.  The average observed core shift is consistent with predictions from the jet model (and with no-jet models) but individual epochs show substantially more scatter.
\label{fig:coreshift}
}
\end{figure}

With the much smaller subset of observations, \citet{Bower15} searched inconclusively for this effect.  With the much larger sample of epochs with two frequency measurements in this paper and excluding the anomalous first four epochs, we find a mean value for the core shift $\Phi=(-0.11 \pm 0.11, -0.02 \pm 0.05)\,{\rm mas\,cm^{-1}}$, that is also consistent with zero.  Different methods of estimation of the mean core shift in Table~\ref{tab:fits} give similar values and uncertainties.  Individual epochs, however, show a large scatter that formally indicates detection in those epochs.  The origin of this scatter is unclear but it is possibly related to the sources of systematic error discussed previously for the overall position errors or to intrinsic  effects in \sgra.  The East-West scatter is significantly larger than than the North-South scatter, consistent with an origin related to the scattering asymmetry.

The measured average meets the expectations for jet models, which predict $\Phi\sim 0.1 {\rm mas\,cm^{-1}}$, comparable to the uncertainty that we achieve with these measurements \citep{2014A&A...570A...7M}.  We show two of these models for inclination angles of 30$^\circ$ and 60$^\circ$ in Figure~\ref{fig:coreshift}.  Constraints based on EHT observations favor models with $i<60^\circ$ \citep{2022ApJ...930L..16E}.  These more recent models based on EHT observations predict core shift of a similar magnitude. It is also possible that the scatter in individual measurements is the result of changes in the large scale structure of \sgra\ as the result of turbulent jet outflows or structures in the accretion flow.  Time variable structures of this kind could have characteristic time scales as short as the variability time scales measured for \sgra\ at these wavelengths, i.e., tens of minutes to hours \citep[e.g.,][]{2015A&A...576A..41B}.

\section{Conclusions}

We have presented extensive new astrometric and high-resolution imaging observations of the GC magnetar \psr, extending until the magnetar flux density weakened to undetectable levels.  The results provide significantly higher accuracy and greater temporal coverage that support previously published conclusions.  The primary results are:
\begin{itemize}
\item The proper motion remains consistent with the population of stars from the CW stellar disk, supporting an origin with modest kick velocity from a star in that disk.
\item Significantly stronger limits are set on the tangential acceleration of \psr, which remain consistent with the expectations for the maximum acceleration due to  \sgra. If the magnetar re-brightens, new observations will have the power to detect the acceleration rapidly given the much longer time baseline.
\item Astrometric residuals in right ascension exhibit an apparently sinusoidal variation with a period $\sim 1600$ days.  Pulse period measurements are, however, inconsistent with a binary companion explanation of the astrometric data.  Alternative possibilities for the variation, regardless of whether or not it is periodic or red-noise, are refractive wander and changes in the structure of \sgra.  None of these models are entirely good fits to all of the data.
\item The stable apparent size of the magnetar over the duration of these observations stands in contrast to the highly variable RM, supporting a picture in which these propagation effects originate from different regions of the ISM.  The results remain consistent with a single turbulent plasma screen at a distance $\sim 5$ kpc away from \sgra.
\item The results are consistent with a minimal stable core shift, as expected from a compact jet or accretion disk.  Individual epochs, however, show significant scatter in the apparent core shift that may be the result of changing structures in \sgra, propagation effects, and/or systematic errors in the calibration and imaging.
\end{itemize}

If \psr\ re-brightens so that it is detectable in astrometric experiments, future measurements can measure the acceleration due to \sgra\ and resolve the question of the presence of a binary companion or other sources of apparent short-term position variation.
This complex set of results from a single, unique GC pulsar highlights the rich science to be gleaned from searches for and characterization of pulsars in this region with existing and future instruments.

\begin{acknowledgements}
The National Radio Astronomy Observatory is a facility of the National Science Foundation operated under cooperative agreement by Associated Universities, Inc. 
We acknowledge financial support by the Italian Ministry of University and Research (MUR)– Project CUP F53D23001260001, funded by the European Union – NextGenerationEU.  
This work was partially supported by FAPESP (Funda\c{c}\~ao de Amparo \'a Pesquisa do Estado de S\~ao Paulo) under grant 2021/01183-8. We acknowledge the European Research
Council (ERC) Synergy Grant “BlackHoleCam: Imaging the Event Horizon of Black Holes" (grant 610058) and Synergy Grant "BlackHolistic:
Colour Movies of Black Holes: Understanding Black Hole Astrophysics from the Event Horizon to Galactic Scales" (grant 10107164).
\end{acknowledgements}

\bibliography{refs}{}
\bibliographystyle{aasjournal}



\end{document}